\documentclass[12pt]{article}
\usepackage{graphicx, amsmath, amssymb}
\usepackage{hyperref}

\newcommand{\mbf}[1]{\mathbf{#1}}

\begin{document}
\begin{flushright}
{SLAC--PUB--13998\\
\date{today}}
\end{flushright}

\vspace{20pt}

\centerline{\LARGE  {AdS/QCD, Light-Front Holography,}}

\vspace{5pt}

\centerline{\LARGE  {and the Nonperturbative Running Coupling}}

\vspace{20pt}

\centerline{ {
Stanley J. Brodsky,$^{a}$
 %
Guy F. de T\'eramond,$^{b}$
%
and
Alexandre Deur$^{c}$
}}

\vspace{20pt}

{\centerline {$^{a}${SLAC National Accelerator Laboratory 
Stanford University, Stanford, CA 94309, USA}}

\vspace{4pt}

{\centerline {$^{b}${Universidad de Costa Rica, San Jos\'e, Costa Rica}}

\vspace{4pt}

{\centerline {$^{c}${Thomas Jefferson National Accelerator Facility,
Newport News, VA 23606, USA}}

 \vspace{30pt}

\centerline{\bf Abstract}

\noindent

\vspace{15pt}

The combination of Anti-de Sitter  space (AdS) methods with light-front (LF) holography provides a remarkably accurate first approximation for the spectra and wavefunctions of meson and baryon light-quark  bound states.  The resulting bound-state Hamiltonian equation of motion in QCD leads to  relativistic light-front wave equations in terms of an invariant impact variable $\zeta$ which measures the separation of the quark and gluonic constituents within the hadron at equal light-front time. These equations of motion in physical space-time are  equivalent to the equations of motion which describe the propagation of spin-$J$ modes in anti--de Sitter (AdS) space. The eigenvalues give the hadronic spectrum, and the eigenmodes represent the probability distributions of the hadronic constituents at a given scale.
A positive-sign confining dilaton background modifying AdS space gives a very good account of meson and baryon spectroscopy and form factors.
The light-front holographic mapping  of this model also leads to a non-perturbative effective coupling
$\alpha_s^{AdS}(Q^2)$ which agrees 
with
the effective charge defined by the Bjorken sum rule and
lattice simulations. It  displays a transition from  perturbative to nonperturbative conformal regimes
at a momentum scale $ \sim 1$ GeV.  The resulting   $\beta$-function  appears to capture the essential characteristics of the full
$\beta$-function of QCD, thus giving further support to the application of the gauge/gravity duality to the confining dynamics of strongly coupled QCD.

\newpage

\section{Introduction}

The AdS/CFT correspondence~\cite{Maldacena:1997re} between a gravity or string theory on a higher dimensional Anti--de Sitter (AdS) space-time with conformal gauge field theories (CFT) in physical space-time has brought  a new set of tools for studying the dynamics of strongly coupled quantum field theories, and it has led to new analytical insights into the confining dynamics of QCD.
The AdS/CFT duality provides a gravity description in a ($d+1$)-dimensional AdS
space-time in terms of a flat
$d$-dimensional conformally-invariant quantum field theory defined at the AdS asymptotic boundary.~\cite{Gubser:1998bc}
Thus, in principle, one can compute physical observables in a strongly coupled gauge theory  in terms of a classical gravity theory.
Since the quantum field theory dual to AdS$_5$ space in the original correspondence~\cite{Maldacena:1997re} is conformal, the strong coupling of the dual gauge theory is constant, and its $\beta$-function is zero. Thus one must consider a deformed AdS space in order to simulate color confinement and have a running coupling $\alpha_s^{AdS}(Q^2)$ for the gauge theory side of the correspondence.   As we shall review here, a positive-sign confining dilaton background modifying AdS space gives a very good account of meson and baryon spectroscopy and their elastic form factors. The light-front holographic mapping  of this model also leads to a non-perturbative effective coupling
$\alpha_s^{AdS}(Q^2)$ which agrees with the effective charge defined by the Bjorken sum rule and
lattice simulations~\cite{new}.

In the standard applications of  AdS/CFT methods, one begins with Maldacena's duality between  the conformal supersymmetric $SO(4,2)$ gauge theory and a semiclassical supergravity string theory defined in a 10 dimension AdS$_5 \times S^5$
space-time.~\cite{Maldacena:1997re} The essential mathematical tool underlying Maldacena's observation is the fact that the effects of scale transformations in a conformal theory can be mapped to the $z$ dependence of amplitudes in AdS$_5$ space.
QCD is not conformal but  there is in fact much empirical evidence from lattice, Dyson Schwinger theory and effective charges that the QCD $\beta$ function vanishes in the infrared.~\cite{Deur:2008rf}  The QCD infrared fixed point arises since the propagators of the confined quarks and gluons in the  loop integrals contributing to the $\beta$ function have a maximal wavelength.~\cite{Brodsky:2008be} The decoupling of quantum loops in the infrared is analogous to QED where vacuum polarization corrections to the photon propagator decouple at $Q^2 \to 0$.

We thus begin with a conformal approximation to QCD to model an effective dual gravity description in AdS space. One uses the  five-dimensional AdS$_5$ geometrical representation of the conformal group to represent scale transformations within the conformal window.  Confinement can be
 introduced with a sharp cut-off in the infrared region of AdS space, as in the ``hard-wall" model,~\cite{Polchinski:2001tt}
 or, more successfully,  using a dilaton background in the fifth dimension to produce a smooth cutoff at large distances 
as  in the ``soft-wall" model.~\cite{Karch:2006pv}
The soft-wall AdS/CFT model with a positive-sign dilaton-modified AdS space leads to the
 potential $U(z) = \kappa^4 z^2 + 2 \kappa^2(L+S-1),$~\cite{deTeramond:2009xk} in the fifth dimension coordinate $z$.
 We assume a dilaton profile $\exp(+\kappa^2 z^2)$~\cite{deTeramond:2009xk, Andreev:2006ct, Zuo:2009dz,Afonin:2010fr}, with  opposite sign  to that of Ref.~\cite{Karch:2006pv}.   The resulting spectrum reproduces linear Regge trajectories, where ${\cal M}^2(S,L,n) $ is proportional to the internal 
 spin,  orbital angular momentum $L$ and the principal quantum number $n$.

The modified metric induced by the dilaton can be interpreted in AdS space as a gravitational potential
for an object of mass $m$  in the fifth dimension:
$V(z) = mc^2 \sqrt{g_{00}} = mc^2 R \, e^{\pm \kappa^2 z^2/2}/z$.
In the case of the negative solution, the potential decreases monotonically, and thus an object in AdS will fall to infinitely large
values of $z$.  For the positive solution, the potential is non-monotonic and has an absolute minimum at $z_0 = 1/\kappa$.
Furthermore, for large values of $z$ the gravitational potential increases exponentially,  confining any object  to distances $\langle z \rangle \sim 1/\kappa$.~\cite {deTeramond:2009xk}  We thus will choose the positive sign dilaton solution. This additional warp factor leads to a well-defined scale-dependent effective coupling.
Introducing a positive-sign dilaton background is also relevant for describing chiral symmetry breaking,~\cite{Zuo:2009dz} since  the expectation value of the scalar field associated with the quark mass and condensate does not blow-up in the far infrared region of AdS, in contrast with the original model.~\cite{Karch:2006pv}

Glazek and Schaden~\cite{Glazek:1987ic} have shown that a  harmonic oscillator confining potential naturally arises as an effective potential between heavy quark states when one stochastically eliminates higher gluonic Fock states. Also, Hoyer~\cite{Hoyer:2009ep} has argued that the Coulomb  and  a linear  potentials are uniquely allowed in the Dirac equation at the classical level. The linear potential  becomes a harmonic oscillator potential in the corresponding Klein-Gordon equation.

Light-front 
(LF) quantization is the ideal framework for  describing the
structure of hadrons in terms of their quark and gluon degrees of
freedom. The 
light-front wavefunctions
(LFWFs) of bound states in QCD are relativistic generalizations of the Schr\"odinger wavefunctions, but they are determined at fixed light-front time $\tau = x^+ = x^0 + x^3$, the time marked by the
front of a light wave \cite{Dirac:1949cp}, rather than at fixed ordinary time
$t.$ They play the same role in
hadron physics that Schr\"odinger wavefunctions play in atomic physics.~\cite{Brodsky:1997de}
The simple structure of the LF vacuum provides an unambiguous
definition of the partonic content of a hadron in QCD.

Light-front  holography \cite{deTeramond:2008ht, Brodsky:2006uqa, Brodsky:2007hb, Brodsky:2008pf, deTeramond:2009xk, deTeramond:2010we} connects
the equations of motion in AdS space and
the Hamiltonian formulation of QCD in physical space-time quantized
on the light front  at fixed LF time.  This correspondence provides a direct connection between the hadronic amplitudes $\Phi(z)$  in AdS space  with  LF wavefunctions $\phi(\zeta)$ describing the quark and gluon constituent structure of hadrons in physical space-time.  
In the case of a meson, $\zeta = \sqrt{x(1-x) {\bf b}^2_\perp}$ is a Lorentz invariant coordinate which measures
the distance between the quark and antiquark; it is analogous to the radial coordinate $r$ in the Schr\"odinger equation.   In effect $\zeta$ represents the off-light-front energy shell or invariant mass dependence of the bound state; it  allows the separation of the dynamics of quark and gluon binding from
the kinematics of constituent spin and internal orbital angular momentum.~\cite{deTeramond:2008ht}
Light-front holography
thus provides a  connection between the description of
hadronic modes in AdS space and the Hamiltonian formulation of QCD in
physical space-time quantized on the light-front  at fixed LF
time $\tau.$

The mapping between the LF invariant variable $\zeta$ and the fifth-dimension AdS coordinate $z$ was originally obtained
by matching the expression for electromagnetic current matrix
elements in AdS space~\cite{Polchinski:2002jw}  with the corresponding expression for the
current matrix element, using LF  theory in physical space
time.~\cite{Brodsky:2006uqa}   It has also been shown that one
obtains the identical holographic mapping using the matrix elements
of the energy-momentum tensor,~\cite{Brodsky:2008pf, Abidin:2008ku} thus  verifying  the  consistency of the holographic
mapping from AdS to physical observables defined on the light front.

The resulting equation for the mesonic $q \bar q$ bound states at fixed light-front time  has the form of a single-variable relativistic Lorentz invariant  Schr\"odinger equation
\begin{equation} \label{eq:QCDLFWE}
\left(-\frac{d^2}{d\zeta^2}
- \frac{1 - 4L^2}{4\zeta^2} + U(\zeta) \right)
\phi(\zeta) = \mathcal{M}^2 \phi(\zeta),
\end{equation}
where the confining potential is $ U(\zeta) = \kappa^4 \zeta^2 + 2 \kappa^2(L+S-1)$
in the soft-wall model.
Its eigenvalues determine the hadronic spectra and its eigenfunctions are related to the light-front wavefunctions
of hadrons for general spin and orbital angular momentum. This
LF wave equation serves as a semiclassical first approximation to QCD,
and it is equivalent to the equations of motion which describe the
propagation of spin-$J$ modes in  AdS space.  The resulting light-front wavefunctions provide a fundamental description of the structure and internal dynamics of hadronic states in terms of their constituent quark and gluons.
There is only one parameter, the mass scale $\kappa \sim 1/2$ GeV, which enters the confinement potential. In the case of mesons $S=0,1$ is the combined spin of the $q $ and $ \bar q $, $L$ is their relative orbital angular momentum as determined by the hadronic light-front wavefunctions.

The concept of a running coupling $\alpha_s(Q^2)$  in QCD is usually restricted to the perturbative domain.  However, as in QED, it is useful to define the coupling as an analytic function valid over the full space-like and time-like domains.
The study of the non-Abelian QCD coupling at small momentum transfer is a complex problem because of  gluonic self-coupling and color confinement.
We will  show that the light-front holographic mapping of classical gravity in AdS space, modified by a positive-sign
dilaton background, leads to a non-perturbative effective coupling
$\alpha_s^{AdS}(Q^2)$ which is in agreement with hadron physics data
extracted from different observables, as well as with  the predictions of models with built-in confinement  and lattice simulations.

\section{The Hadron Spectrum 
and Form Factors
in Light-Front AdS/QCD}

The meson spectrum predicted by  Eq. \ref{eq:QCDLFWE} has a string-theory Regge form
${\cal M}^2 = 4 \kappa^2(n+ L+S/2)$; {\it i.e.}, the square of the eigenmasses are linear in both $L$ and $n$, where $n$ counts the number of nodes  of the wavefunction in the radial variable $\zeta$.  This is illustrated for the pseudoscalar and vector meson spectra in Fig. \ref{pionVM},
where the data are from Ref.~\cite{Amsler:2008xx}.
The pion ($S=0, n=0, L=0$) is massless for zero quark mass, consistent with the chiral invariance of massless QCD.  Thus one can compute the hadron spectrum by simply adding  $4 \kappa^2$ for a unit change in the radial quantum number, $4 \kappa^2$ for a change in one unit in the orbital quantum number  $L$ and $2 \kappa^2$ for a change of one unit of spin $S$. Remarkably, the same rule holds for three-quark baryons as we shall show below.

\begin{figure}[h]
\begin{center}
\includegraphics[width=6.8cm]{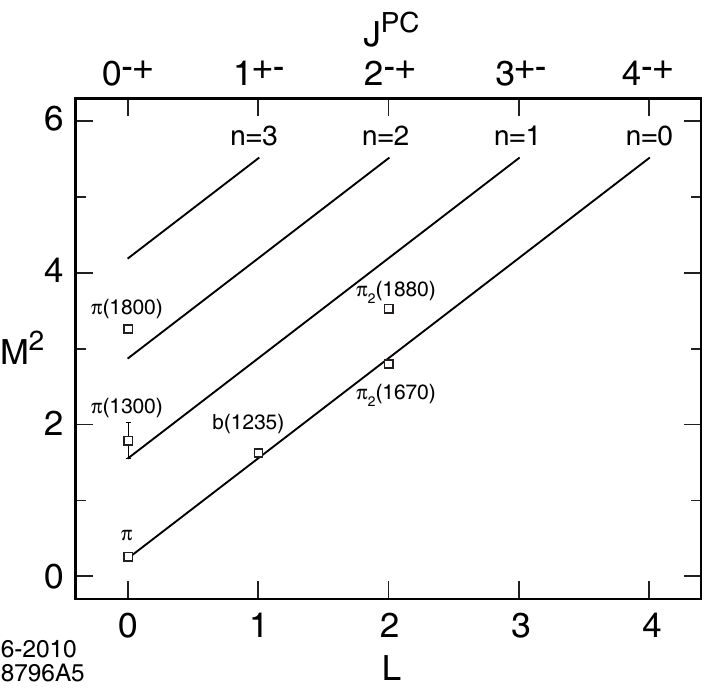}  ~~~
\includegraphics[width=6.8cm]{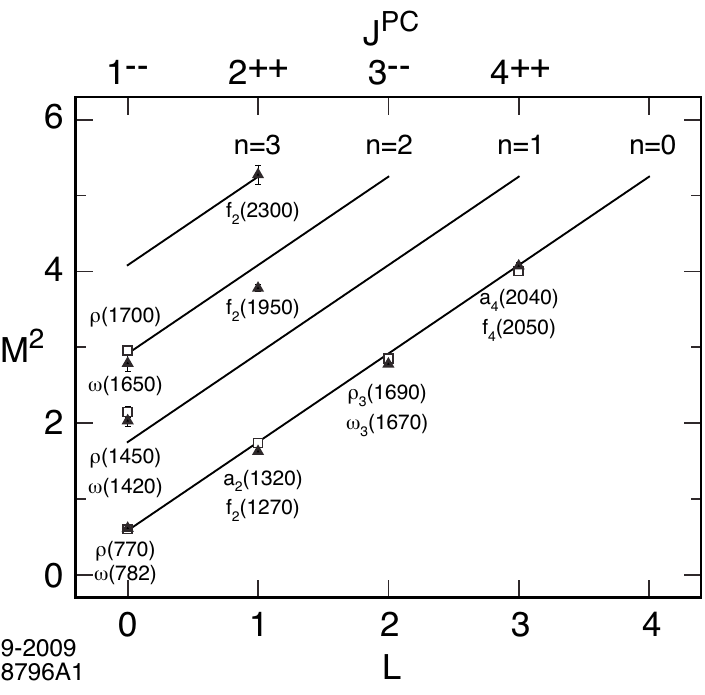}
 \caption{Parent and daughter Regge trajectories for (a) the $\pi$-meson family with
$\kappa= 0.6$ GeV; and (b) the  $I\!=\!1$ $\rho$-meson
 and $I\!=\!0$  $\omega$-meson families with $\kappa= 0.54$ GeV}
\label{pionVM}
\end{center}
\end{figure}

In the light-front formalism, one  sets boundary conditions at fixed $\tau$ and then evolves the system using the light-front (LF) Hamiltonian $P^-  \!= \!
P^0-P^3 = i {d/d \tau}$.  The invariant Hamiltonian $H_{LF} = P^+ P^- \! - P^2_\perp$ then has eigenvalues $\mathcal{M}^2$ where $\mathcal{M}$
is the physical mass.   Its eigenfunctions are the light-front eigenstates whose Fock state projections define the frame-independent light-front wavefunctions.
The eigensolutions of  Eq. \ref{eq:QCDLFWE} provide the light-front wavefunctions of the valence Fock state of the hadrons $\psi(x,
\bf{b}_\perp)$  as illustrated for the pion in Fig. \ref{LFWFPionFFSL} for the soft (a) and hard wall (b) models.   The resulting distribution amplitude has a
broad form $\phi_\pi(x) \sim \sqrt{x(1-x)}$ which is compatible with moments determined from lattice gauge theory. One can then immediately
compute observables such as hadronic form factors (overlaps of LFWFs), structure functions (squares of LFWFs), as well as the generalized parton
distributions and distribution amplitudes which underly hard exclusive reactions. For example, hadronic form factors can be predicted from the
overlap of LFWFs in the Drell-Yan West formula. The prediction for the space-like pion form factor is shown in Fig.  \ref{LFWFPionFFSL}(c). The
pion form factor and the vector meson poles residing in the dressed current in the soft wall model require choosing  a value of $\kappa$ smaller
by a factor of $1/\sqrt 2$  than the canonical value of  $\kappa$ which determines the mass scale of the hadronic spectra.  This shift is
apparently due to the fact that the transverse current in $e^+ e^- \to q \bar q$ creates a quark pair with $L^z= \pm 1$ instead of the $L^z=0$
$q \bar q$ composition of the vector mesons in the spectrum.

\begin{figure}[h]
\begin{center}
 \includegraphics[width=8.0cm]{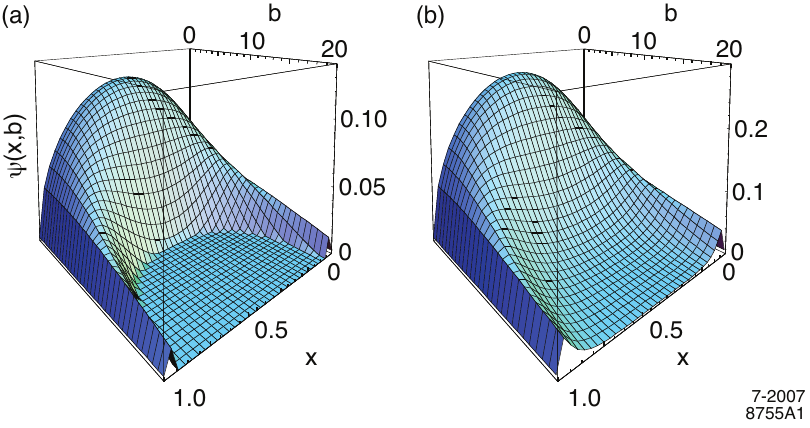}~~
\includegraphics[width=5.0cm]{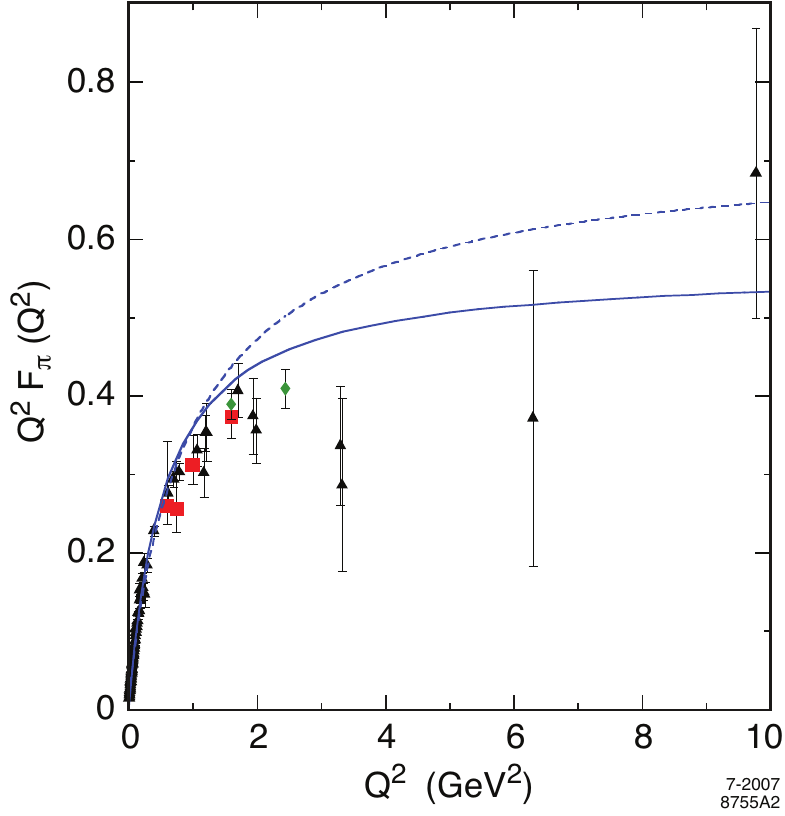}
 \caption{ Pion LF wavefunction $\psi_\pi(x, \bf{b}_\perp$) for the  AdS/QCD (a) hard-wall ($\Lambda_{QCD} = 0.32$ GeV) and (b) soft-wall  ($\kappa = 0.375$ GeV)  models.  (c) Space-like scaling behavior for $Q^2 F_\pi(Q^2).$ The continuous line is the prediction of the soft-wall model for
$\kappa = 0.375$ GeV.  The dashed line is the prediction of the hard-wall model for $\Lambda_{\rm QCD} = 0.22$ GeV. The triangles are the data
compilation of Baldini.}
\label{LFWFPionFFSL}
\end{center}
\end{figure}

Individual hadrons in AdS/QCD are identified by matching the power behavior of the hadronic amplitude at the AdS boundary at small $z$ to the twist $\tau$ of its interpolating operator at short distances $x^2 \to 0$, as required by the AdS/CFT dictionary. The twist also equals the dimension of fields appearing in chiral super-multiplets;~\cite{Craig:2009rk}
thus the twist of a hadron equals the number of constituents plus the relative orbital angular momentum.
One then can apply light-front holography to relate the amplitude eigensolutions  in the fifth dimension coordinate $z$  to the LF wavefunctions in the physical space-time variable  $\zeta$.

Equation (\ref{eq:QCDLFWE}) was derived by taking the LF bound-state Hamiltonian equation of motion as the starting
point.~\cite{deTeramond:2008ht} The term $L^2/ \zeta^2$  in the  LF equation of motion  (\ref{eq:QCDLFWE})
is derived from  the reduction of the LF kinetic energy when one transforms 
to two-dimensional cylindrical coordinates $(\zeta, \varphi)$,
in analogy to the $\ell(\ell+1)/ r^2$ Casimir term in Schr\"odinger theory.  One thus establishes the interpretation of $L$ in the AdS equations of motion.
The interaction terms build confinement  corresponding to
the dilaton modification of AdS space~\cite{deTeramond:2008ht}.
The duality between these two methods provides a direct
connection between the description of hadronic modes in AdS space and
the Hamiltonian formulation of QCD in physical space-time quantized
on the light-front  at fixed LF time $\tau.$

The identification of orbital angular momentum of the constituents is a key element in the description of the internal structure of hadrons using holographic principles. In our approach  quark and gluon degrees of freedom are explicitly introduced in the gauge/gravity correspondence,~\cite{Brodsky:2003px} in contrast with the usual
AdS/QCD framework~\cite{Erlich:2005qh,DaRold:2005zs} where axial and vector currents become the primary entities as in effective chiral theory.
Unlike the top-down string theory approach,  one is not limited to hadrons of maximum spin
$J \le 2$, and one can study baryons with finite color $N_C=3.$   Higher spin modes follow from shifting dimensions in the AdS wave equations.
In the soft-wall
model the usual Regge behavior is found $\mathcal{M}^2 \sim n +
L$, predicting the same multiplicity of states for mesons
and baryons as observed experimentally.~\cite{Klempt:2007cp}
It is possible to extend the model to hadrons with heavy quark constituents
by introducing nonzero quark masses and short-range Coulomb
corrections.  For other
recent calculations of the hadronic spectrum based on AdS/QCD, see Refs.~\cite{BoschiFilho:2005yh, Evans:2006ea, Hong:2006ta, Colangelo:2007pt, Forkel:2007ru, Vega:2008af, Nawa:2008xr, dePaula:2008fp,  Colangelo:2008us, Forkel:2008un, Ahn:2009px, Sui:2009xe}. Other recent computations of the pion form factor are given in
Refs.~\cite{Kwee:2007dd, Grigoryan:2007wn}.

For baryons, the light-front wave equation is a linear equation
determined by the LF transformation properties of spin 1/2 states. A linear confining potential
$U(\zeta) \sim \kappa^2 \zeta$ in the LF Dirac
equation leads to linear Regge trajectories.~\cite{Brodsky:2008pg}   For fermionic modes the light-front matrix
Hamiltonian eigenvalue equation $D_{LF} \vert \psi \rangle = \mathcal{M} \vert \psi \rangle$, $H_{LF} = D_{LF}^2$,
in a $2 \times 2$ spinor  component
representation is equivalent to the system of coupled linear equations
\begin{eqnarray} \label{eq:LFDirac} \nonumber
- \frac{d}{d\zeta} \psi_- -\frac{\nu+{1\over 2}}{\zeta}\psi_-
- \kappa^2 \zeta \psi_-&=&
\mathcal{M} \psi_+, \\ \label{eq:cD2k}
  \frac{d}{d\zeta} \psi_+ -\frac{\nu+{1\over 2}}{\zeta}\psi_+
- \kappa^2 \zeta \psi_+ &=&
\mathcal{M} \psi_-,
\end{eqnarray}
with eigenfunctions
\begin{eqnarray} \nonumber
\psi_+(\zeta) &\sim& z^{\frac{1}{2} + \nu} e^{-\kappa^2 \zeta^2/2}
  L_n^\nu(\kappa^2 \zeta^2) ,\\
\psi_-(\zeta) &\sim&  z^{\frac{3}{2} + \nu} e^{-\kappa^2 \zeta^2/2}
 L_n^{\nu+1}(\kappa^2 \zeta^2),
\end{eqnarray}
and  eigenvalues $\mathcal{M}^2 = 4 \kappa^2 (n + \nu + 1)$.

\begin{figure}[h]
\begin{center}
\includegraphics[width=13.6cm]{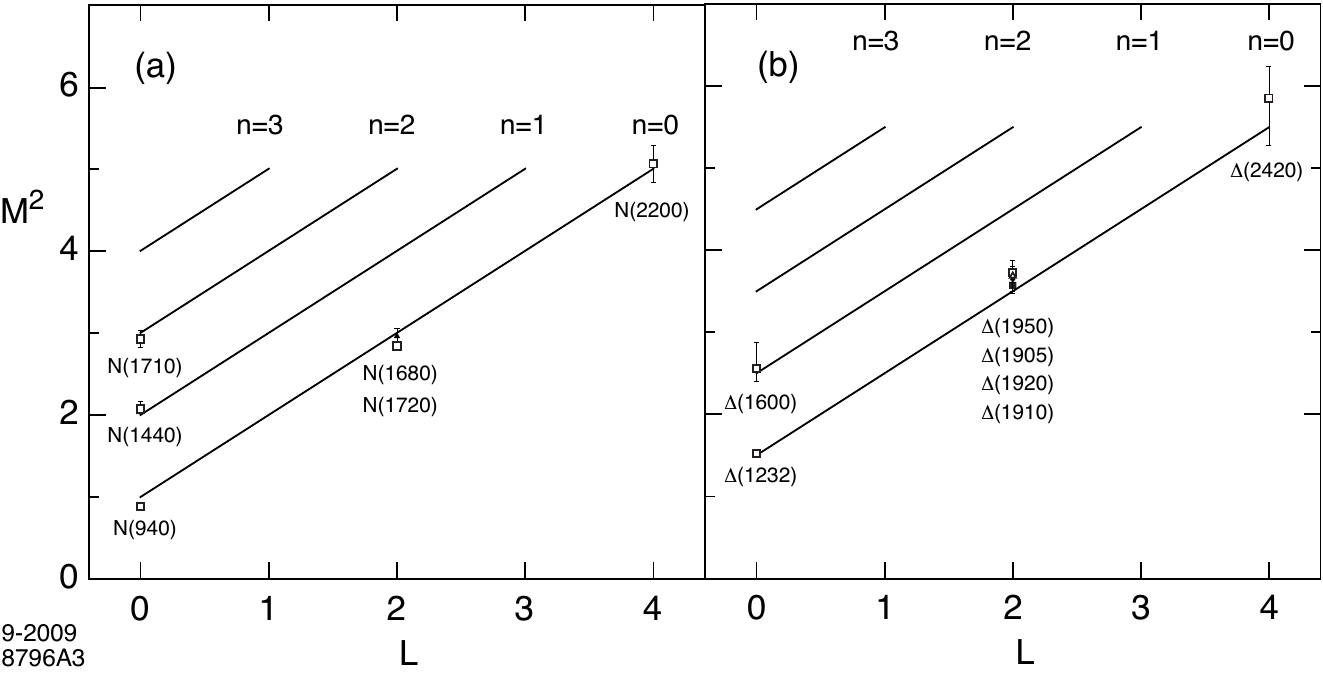}
\caption{{\bf 56} Regge trajectories for  the  $N$ and $\Delta$ baryon families for $\kappa= 0.5$ GeV.}
\label{Baryons}
\end{center}
\end{figure}

The baryon interpolating operator
$ \mathcal{O}_{3 + L} =  \psi D_{\{\ell_1} \dots
 D_{\ell_q } \psi D_{\ell_{q+1}} \dots
 D_{\ell_m\}} \psi$,  $L = \sum_{i=1}^m \ell_i$, is a twist 3,  dimension $9/2 + L$ with scaling behavior given by its
 twist-dimension $3 + L$. We thus require $\nu = L+1$ to match the short distance scaling behavior. Higher spin modes are obtained by shifting dimensions for the fields.
 Thus, as in the meson sector,  the increase  in the
mass squared for higher baryonic state is
$\Delta n = 4 \kappa^2$, $\Delta L = 4 \kappa^2$ and $\Delta S = 2 \kappa^2,$
relative to the lowest ground state,  the proton. Since our starting point to find the bound state equation of motion for baryons is the light-front, we fix the overall energy scale identical for mesons and baryons by imposing chiral symmetry to the pion~\cite{deTeramond:2010we} in the LF Hamiltonian equations. By contrast, if we start with a five-dimensional action for a scalar field in presence of a positive sign dilaton, the pion is automatically massless.

The predictions for the $\bf 56$-plet of light baryons under the $SU(6)$  flavor group are shown in Fig. \ref{Baryons}.
As for the predictions for mesons in Fig. \ref{pionVM}, only confirmed PDG~\cite{Amsler:2008xx} states are shown.
The Roper state $N(1440)$ and the $N(1710)$ are well accounted for in this model as the first  and second radial
states. Likewise the $\Delta(1660)$ corresponds to the first radial state of the $\Delta$ family. The model is  successful in explaining the important parity degeneracy observed in the light baryon spectrum, such as the $L\! =\!2$, $N(1680)\!-\!N(1720)$ degenerate pair and the $L=2$, $\Delta(1905), \Delta(1910), \Delta(1920), \Delta(1950)$ states which are degenerate
within error bars. Parity degeneracy of baryons is also a property of the hard wall model, but radial states are not well described in this model.~\cite{deTeramond:2005su}

As an example  of the scaling behavior of a twist $\tau = 3$ hadron, we compute the spin non-flip
nucleon form factor in the soft wall model.~\cite{Brodsky:2008pg} The proton and neutron Dirac
form factors are given by
\begin{equation}
F_1^p(Q^2) =  \! \int  d \zeta \, J(Q, \zeta) \,
  \vert \psi_+(\zeta)\vert^2 ,
\end{equation}
\begin{equation}
F_1^n(Q^2) =  - \frac{1}{3}  \! \int  d \zeta  \,  J(Q, \zeta)
 \left[\vert \psi_+(\zeta)\vert^2 - \vert\psi_-(\zeta)\vert^2\right],
 \end{equation}
where $F_1^p(0) = 1$,~ $F_1^n(0) = 0$. The non-normalizable mode
 $J(Q,z)$ is the solution of the
AdS wave equation for the external electromagnetic current in presence of a dilaton
background  $\exp(\pm \kappa^2 z^2)$.~\cite{Brodsky:2007hb, Grigoryan:2007my}
Plus and minus components of the twist 3 nucleon LFWF are
\begin{equation} \label{eq:PhipiSW}
\psi_+(\zeta) \!=\! \sqrt{2} \kappa^2 \, \zeta^{3/2}  e^{-\kappa^2 \zeta^2/2},  ~~~
\Psi_-(\zeta) \!=\!  \kappa^3 \, \zeta^{5/2}  e^{-\kappa^2 \zeta^2/2}.
\end{equation}
The results for $Q^4 F_1^p(Q^2)$ and $Q^4 F_1^n(Q^2)$   and are shown in
Fig. \ref{fig:nucleonFF}.~\cite{note3}

\begin{figure}[h]
\begin{center}
 \includegraphics[width=7.6cm]{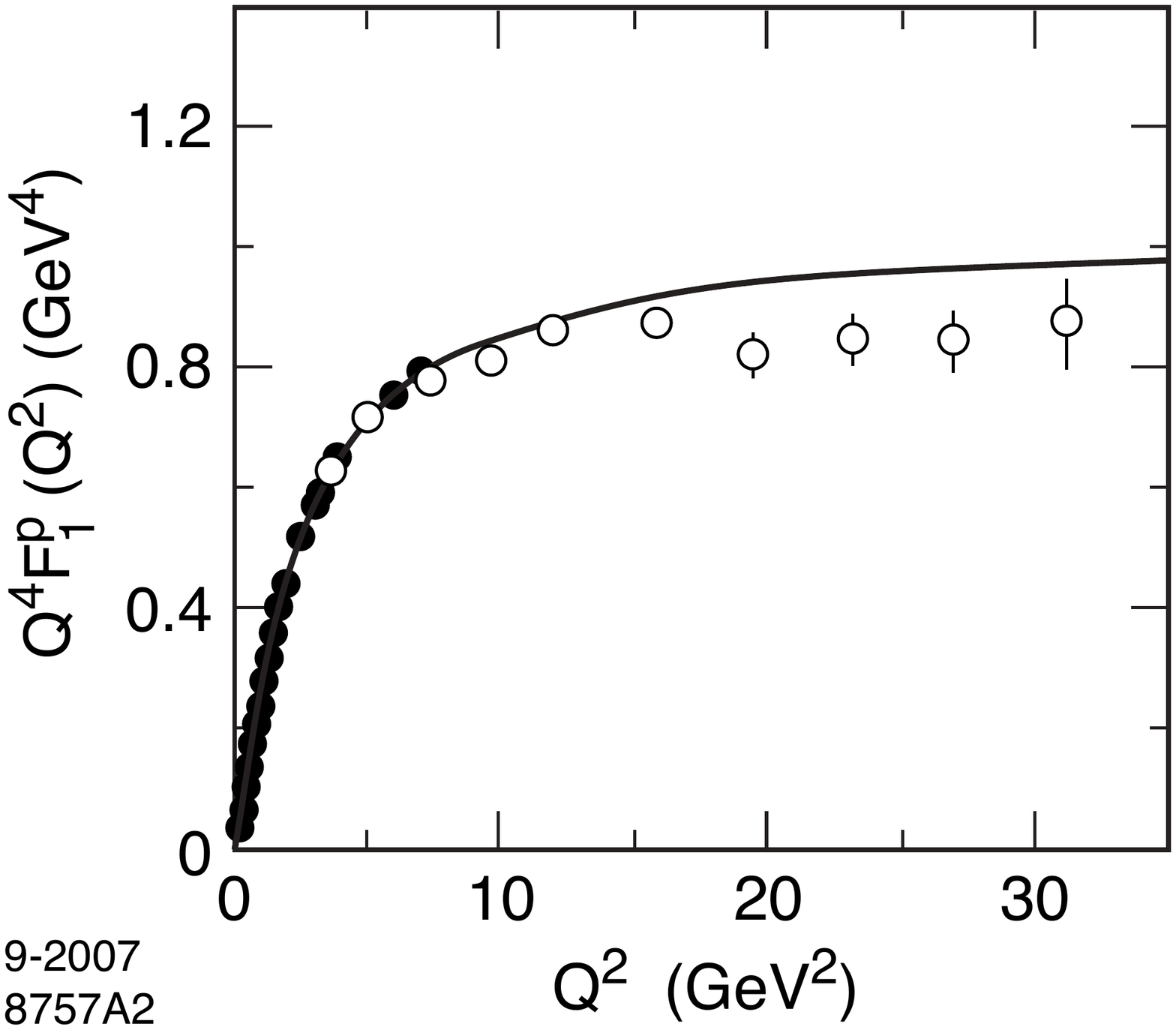}
\includegraphics[width=7.4cm]{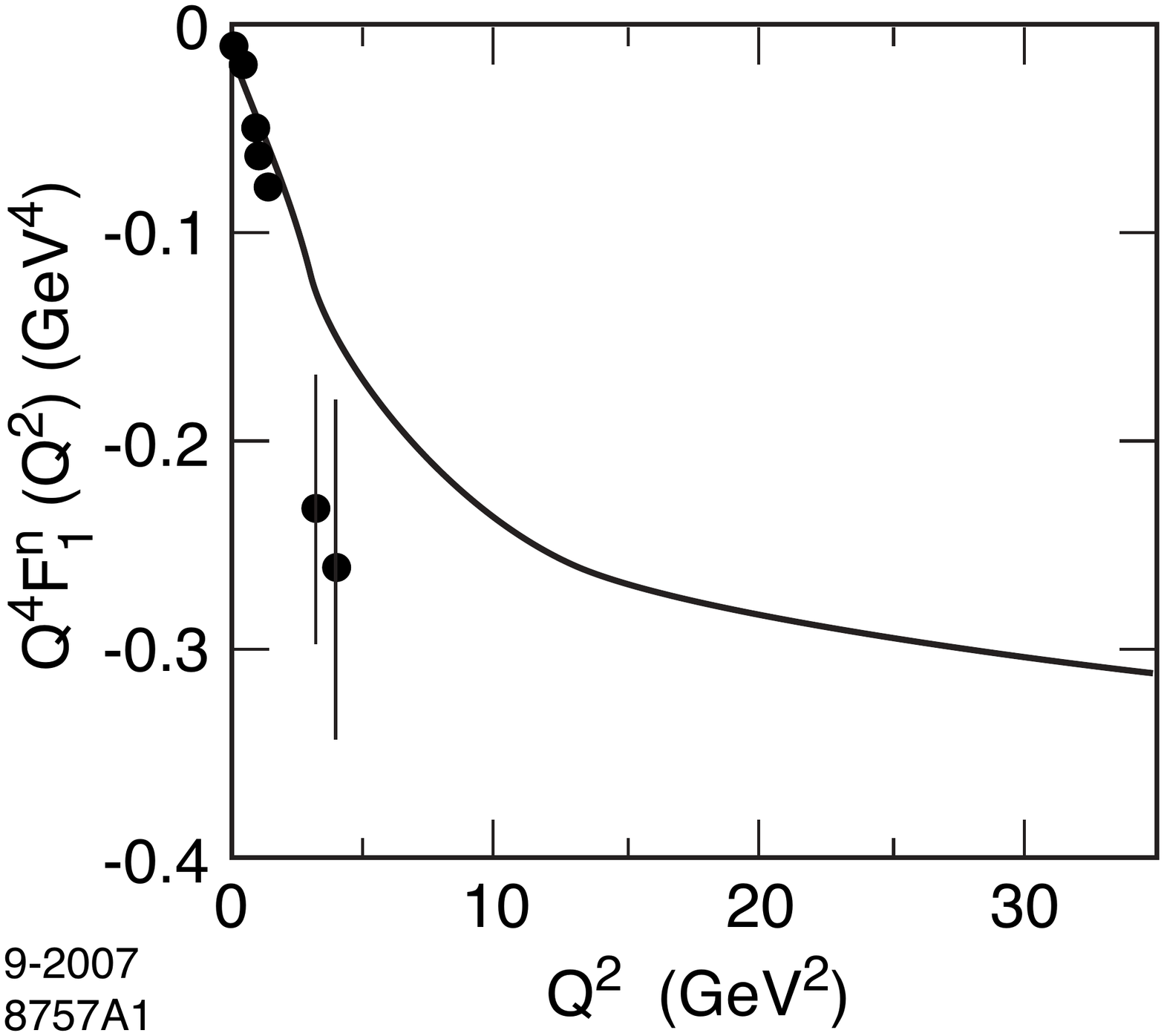}
 \caption{Predictions for $Q^4 F_1^p(Q^2)$ and $Q^4 F_1^n(Q^2)$ in the
soft wall model for $\kappa =  0.424$ GeV.}
\label{fig:nucleonFF}
\end{center}
\end{figure}

\section{Nonperturbative Running Coupling from Light-Front Holography \label{alphaAdS}}

The definition of the running coupling in perturbative quantum field theory is scheme-dependent.  As discussed by Grunberg,~\cite{Grunberg}  an effective coupling or charge can be defined directly from physical observables.
Effective charges defined from
different observables can be related  to each other in the leading-twist domain using commensurate scale relations
 (CSR).~\cite{CSR}    The  potential between infinitely heavy quarks can be defined analytically in momentum transfer
 space as the product  of the running coupling times the Born gluon propagator: $V(q)  = - 4 \pi C_F {\alpha_V(q) / q^2}$.   This effective charge defines a renormalization scheme -- the $\alpha_V$ scheme of Appelquist, Dine,  and Muzinich.~\cite{Appelquist:1977tw}
In fact, the holographic coupling $\alpha_s^{AdS}(Q^2)$ can be considered to be the nonperturbative extension of the
$\alpha_V$ effective charge defined in Ref. \cite{Appelquist:1977tw}.
We can also make  use of the $g_1$ scheme, where the strong coupling $\alpha_{g_1}(Q^2)$ is determined from
the Bjorken sum rule.~\cite{BjorkenSR}  The coupling $\alpha_{g_1}(Q^2)$ has the advantage that it is the best-measured effective charge, and it can be used to extrapolate the definition of the effective coupling to large distances.~\cite{Deur:2009hu}  Since $\alpha_{g_1}$ has been measured at intermediate energies, it is 
particularly useful for studying  the transition from 
small
to large distances.

We will show~\cite{new}   how the LF holographic mapping of effective classical gravity in AdS space, modified by a positive-sign dilaton background, can  be used to identify an analytically simple  color-confining
non-perturbative effective coupling $\alpha_s^{AdS}(Q^2)$ as a function of the space-like momentum transfer $Q^2 = - q^2$.   This coupling incorporates  confinement
and agrees well with effective charge observables and lattice simulations.
It also exhibits an infrared fixed point at small $Q^2$ and asymptotic freedom at large $Q^2$. However, the fall-off   of
$\alpha_s^{AdS}(Q^2)$  at large $Q^2$ is exponential: $\alpha_s^{AdS}(Q^2) \sim e^{-Q^2 /  \kappa^2}$, rather than the perturbative QCD (pQCD) logarithmic fall-off.   We also show in Ref. \cite{new} that a phenomenological extended coupling can be defined which implements the pQCD behavior.

As will be discussed below, the  $\beta$-function derived from light-front holography becomes significantly  negative in the non-perturbative regime $Q^2 \sim \kappa^2$, where it reaches a minimum, signaling the transition region from the infrared (IR) conformal region, characterized by hadronic degrees of freedom,  to a pQCD conformal ultraviolet (UV)  regime where the relevant degrees of freedom are the quark and gluon constituents.  The  $\beta$-function is always negative:  it vanishes at large $Q^2$ consistent with asymptotic freedom, and it vanishes at small $Q^2$ consistent with an infrared fixed point.~\cite{Cornwall:1981zr, Brodsky:2008be}

Let us consider a five-dimensional gauge field $G$ propagating in AdS$_5$ space in presence of a dilaton background
$\varphi(z)$ which introduces the energy scale $\kappa$ in the five-dimensional action.
At quadratic order in the field strength the action is
\begin{equation}
S =  - {1\over 4}\int \! d^5x \, \sqrt{g} \, e^{\varphi(z)}  {1\over g^2_5} \, G^2,
\label{eq:action}
\end{equation}
where the metric determinant of AdS$_5$ is $\sqrt g = ( {R/z})^5$,  $\varphi=  \kappa^2 z^2$ and the square of the coupling $g_5$ has dimensions of length.   We  can identify the prefactor
\begin{equation} \label{eq:flow}
g^{-2}_5(z) =  e^{\varphi(z)}  g^{-2}_5 ,
\end{equation}
in the  AdS  action (\ref{eq:action})  as the effective coupling of the theory at the length scale $z$.
The coupling $g_5(z)$ then incorporates the non-conformal dynamics of confinement. The five-dimensional coupling $g_5(z)$
is mapped,  modulo a  constant, into the Yang-Mills (YM) coupling $g_{YM}$ of the confining theory in physical space-time using light-front holography. One  identifies $z$ with the invariant impact separation variable $\zeta$ which appears in the LF Hamiltonian:
$g_5(z) \to g_{YM}(\zeta)$. Thus
\begin{equation}  \label{eq:gYM}
\alpha_s^{AdS}(\zeta) = g_{YM}^2(\zeta)/4 \pi \propto  e^{-\kappa^2 \zeta^2} .
\end{equation}

In contrast with the 3-dimensional radial coordinates of the non-relativistic Schr\"odinger theory, the natural light-front
variables are the two-dimensional cylindrical coordinates $(\zeta, \phi)$ and the
light-cone fraction $x$. The physical coupling measured at the scale $Q$ is the two-dimensional Fourier transform
of the  LF transverse coupling $\alpha_s^{AdS}(\zeta)$  (\ref{eq:gYM}). Integration over the azimuthal angle
 $\phi$ gives the Bessel transform
 \begin{equation} \label{eq:2dimFT}
\alpha_s^{AdS}(Q^2) \sim \int^\infty_0 \! \zeta d\zeta \,  J_0(\zeta Q) \, \alpha_s^{AdS}(\zeta),
\end{equation}
in the $q^+ = 0$ light-front frame where $Q^2 = -q^2 = - \mbf{q}_\perp^2 > 0$ is the square of the space-like
four-momentum transferred to the
hadronic bound state.   Using this ansatz we then have from  Eq.  (\ref{eq:2dimFT})
\begin{equation}
\label{eq:alphaAdS}
\alpha_s^{AdS}(Q^2) = \alpha_s^{AdS}(0) \, e^{- Q^2 /4 \kappa^2}.
\end{equation}
In contrast, the negative dilaton solution $\varphi=  -\kappa^2 z^2$ leads to an integral which diverges at large $\zeta$.
We identify $\alpha_s^{AdS}(Q^2)$ with the physical QCD running coupling
in its nonperturbative domain.

The flow equation  (\ref{eq:flow}) from the scale dependent measure for the gauge fields can be understood as a consequence of field-strength renormalization.
In physical QCD we can rescale the non-Abelian gluon field  $A^\mu \to \lambda A^\mu$  and field strength
$G^{\mu \nu}  \to \lambda G^{\mu \nu}$  in the QCD Lagrangian density  $\mathcal{L}_{\rm QCD}$ by a compensating rescaling of the coupling strength $g \to \lambda^{-1} g.$  The renormalization of the coupling $g _{phys} = Z^{1/2}_3  g_0,$   where $g_0$ is the bare coupling in the Lagrangian in the UV-regulated theory,  is thus  equivalent to the renormalization of the vector potential and field strength: $A^\mu_{ren} =  Z_3^{-1/2} A^\mu_0$, $G^{\mu \nu}_{ren} =  Z_3^{-1/2} G^{\mu \nu}_0$   with a rescaled Lagrangian density
${\cal L}_{\rm QCD}^{ren}  = Z_3^{-1}  { \cal L}_{\rm QCD}^0  = (g_{phys}/g_0)^{-2}  \mathcal{L}_0$.
 In lattice gauge theory,  the lattice spacing $a$ serves as the UV regulator, and the renormalized QCD coupling is determined  from the normalization of the gluon field strength as it appears  in the gluon propagator. The inverse of the lattice size $L$ sets the mass scale of the resulting running coupling.
As is the case in lattice gauge theory, color confinement in AdS/QCD reflects nonperturbative dynamics at large distances. The QCD couplings defined from lattice gauge theory and the soft wall holographic model are thus similar in concept, and both schemes are expected to have similar properties in the nonperturbative domain, up to a rescaling of their respective momentum scales.

\section{Comparison of  the Holographic Coupling with Other Effective Charges \label{alphatest}}

The effective coupling  $\alpha^{AdS}(Q^2)$ (solid line) is compared in Fig. \ref{alphas} with  experimental and lattice data. For this comparison to be meaningful, we have to impose the same normalization on the AdS coupling as the $g_1$ coupling. This defines $\alpha_s^{AdS}$ normalized to the $g_1$ scheme: $\alpha_{g_1}^{AdS}\left(Q^2 \! =0\right) = \pi.$
Details on the comparison with other effective charges are given in Ref. ~\cite{Deur:2005cf}.

\begin{figure}[h]
\begin{center}
\includegraphics[width=7.0cm]{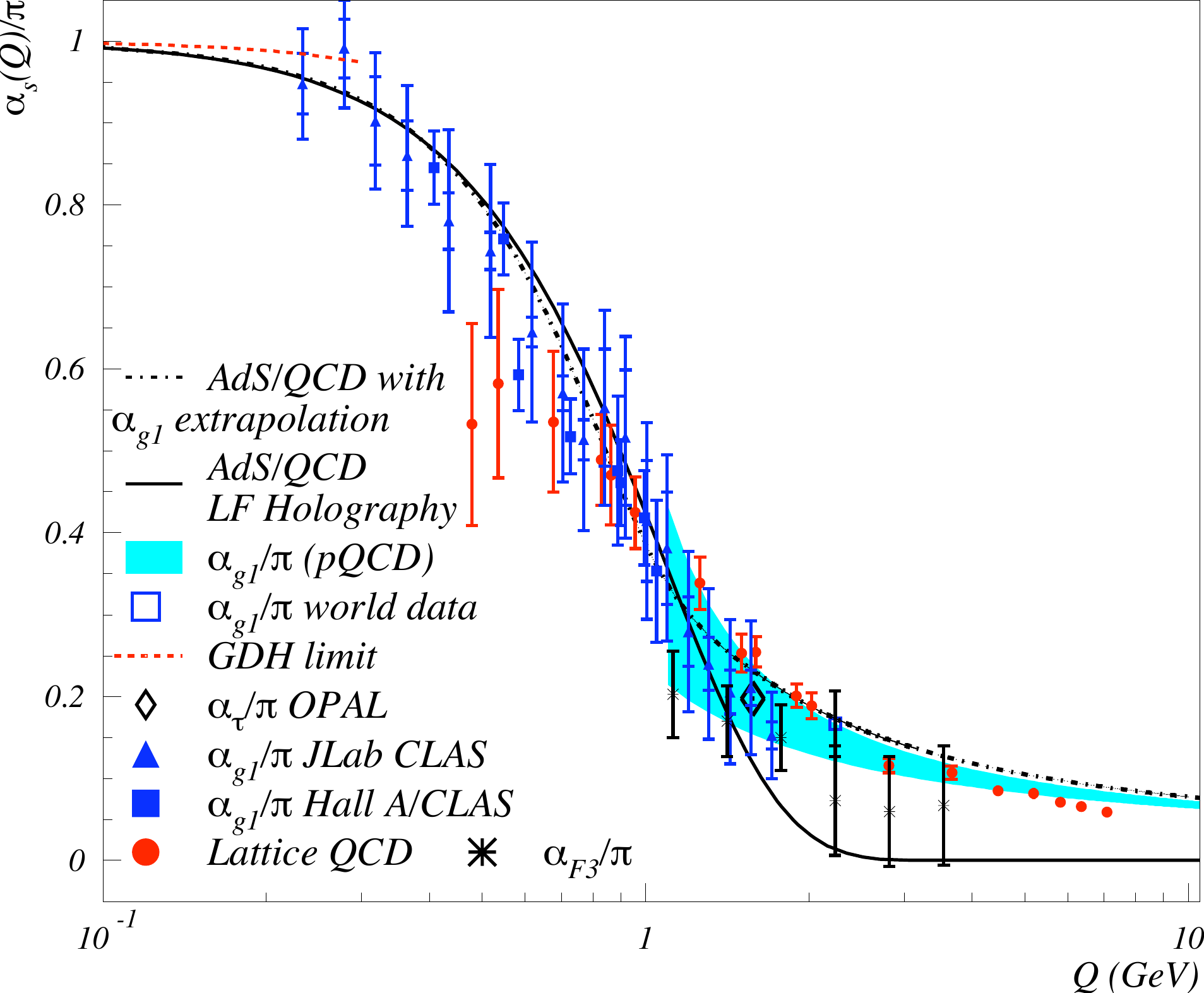} ~~~
\includegraphics[width=6.8cm]{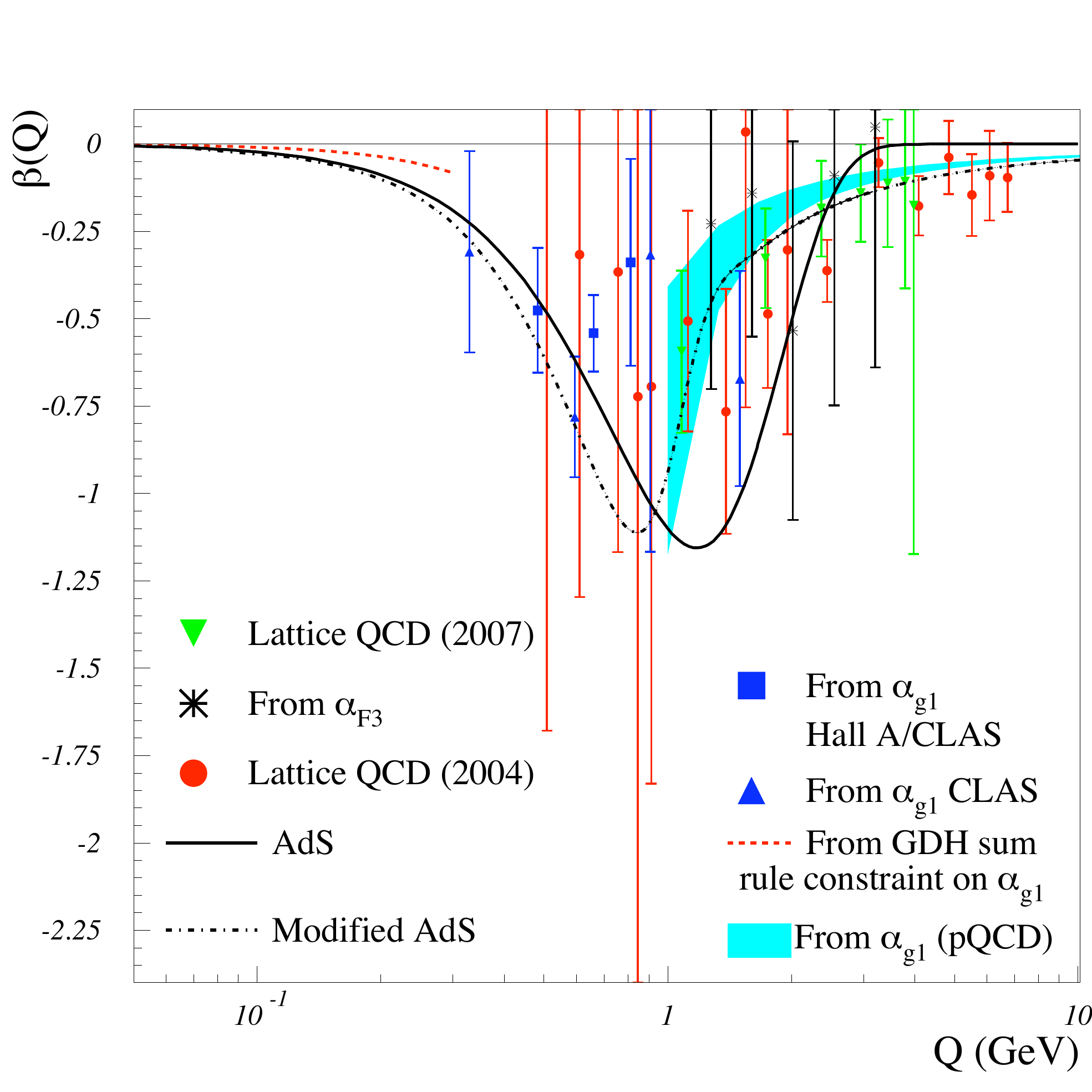}
 \caption{(a) Effective coupling  from LF holography  for  $\kappa = 0.54 ~ {\rm GeV}$ compared with effective QCD couplings  extracted from
different observables and lattice results. (b) Prediction for the $\beta$ function compared to  lattice simulations,  JLab and CCFR results  for the Bjorken sum rule effective charge.}
\label{alphas}
\end{center}
\end{figure}

The couplings in Fig. \ref{alphas} (a) agree well in the strong coupling regime  up to $Q  \! \sim \! 1$ GeV.  The value $\kappa  = 0.54 ~ {\rm GeV}$ is determined from the vector meson 
Regge trajectory.~\cite {deTeramond:2009xk}.
The lattice results shown in Fig. \ref{alphas} from Ref.~\cite{Furui} have been scaled to match the perturbative UV domain. The effective charge $\alpha_{ g_1}$ has been determined in Ref.~\cite{Deur:2005cf} from several experiments.  Fig. \ref{alphas} also displays other couplings from different observables as well as $\alpha_{g_1}$ which is computed from the
Bjorken sum rule~\cite{BjorkenSR}  over a large range of momentum transfer (cyan band). At $Q^2\!=\!0$ one has the constraint  on the slope of $\alpha_{g_1}$ from the Gerasimov-Drell-Hearn (GDH) sum rule~\cite{GDH} which is also shown in the figure.
The results show no sign of a phase transition, cusp, or other non-analytical behavior, a fact which allows us to extend the functional dependence of the coupling to large distances. As discussed below, the smooth behavior of the  AdS strong 
coupling also allows us to extrapolate its form to the perturbative domain.

The hadronic model obtained from the dilaton-modified AdS space provides a semi-classical first approximation to QCD.  Color confinement is introduced by the harmonic oscillator potential, but effects from gluon creation and absorption are not included in this effective theory.  The nonperturbative  confining effects vanish exponentially at large momentum transfer
(Eq. (\ref{eq:alphaAdS})), and thus the logarithmic fall-off from pQCD quantum loops will dominate in this regime.

The  running coupling  $\alpha_s^{AdS}$ given by Eq.  (\ref{eq:alphaAdS})  is obtained from a
color-confining potential.   Since the strong coupling is an analytical function of the momentum transfer at all scales, we can extend  the range of applicability of $\alpha_s^{AdS}$ by matching to a  perturbative coupling
at the transition scale, $Q \sim 1$ GeV, where pQCD contributions become important.
In order to have a fully analytical model, we write
\begin{equation}
\label{eq:alphafit}
\alpha_{Modified, g_1}^{AdS}(Q^2) = \alpha_{g_1}^{AdS}(Q^2) g_+(Q^2 ) + \alpha_{g_1}^{fit}(Q^2) g_-(Q^2),
\end{equation}
where $g_{\pm}(Q^2) = 1/(1+e^{\pm \left(Q^2 - Q^2_0\right)/\tau^2})$ are smeared step functions which match the two
regimes. The parameter $\tau$ represents the width of the
transition region.    Here $\alpha_{g_1}^{AdS}$ is given by Eq. (\ref{eq:alphaAdS}) with the normalization  $\alpha_{g_1}^{AdS}(0)=\pi$ 
-- the plain black line in Fig.~\ref{alphas} -- and $\alpha_{g_{1}}^{fit}$ in Eq. (\ref{eq:alphafit}) is the analytical
fit to the measured coupling $\alpha_{g_1}$.~\cite{Deur:2005cf}
The couplings  are chosen to have the same  normalization at $Q^2=0.$
The smoothly extrapolated result (dot-dashed line) for  $\alpha_{s}$ is also shown on Fig.~\ref{alphas}. We use the
parameters $Q_{0}^{2}=0.8$ GeV$^{2}$ and $\tau^2=0.3$ GeV$^{2}$.

The $\beta$-function for the nonperturbative  effective coupling obtained from the LF holographic mapping in a positive dilaton modified AdS background  is
\begin{equation} \label{eq:beta}
\beta^{AdS}(Q^2)  = {d \over d \log{Q^2}}\alpha^{AdS}(Q^2) = {\pi Q^2\over 4 \kappa^2} e^{-Q^2/(4 \kappa^2)}.
\end{equation}
The solid line in Fig. \ref{alphas} (b) corresponds to the light-front holographic result Eq.  (\ref{eq:beta}).    Near $Q_0 \simeq 2 \kappa \simeq 1$ GeV, we can interpret the results as a transition from  the nonperturbative IR domain to the quark and gluon degrees of freedom in the perturbative UV  regime. The transition momentum  scale $Q_0$  is compatible with the momentum transfer for the onset of scaling behavior in exclusive reactions where quark counting rules are observed.~\cite{Brodsky:1973kr}
For example, in deuteron photo-disintegration the onset of scaling corresponds to  momentum transfer  of  1.0  GeV to the nucleon involved.~\cite{Gao:2004zh}  Dimensional counting is built into the AdS/QCD soft and hard wall models since the AdS amplitudes $\Phi(z)$ are governed by their twist scaling behavior $z^\tau$ at short distances, $ z \to 0$.~\cite{Polchinski:2001tt}

Also shown on Fig. \ref{alphas} (b) are the $\beta$-functions obtained from phenomenology and lattice calculations. For clarity, we present only the LF holographic predictions, the lattice results from,  \cite{Furui} and the
experimental data supplemented by the relevant sum rules.
The width of the aqua band is computed from the uncertainty of $\alpha_{g_1}$
in the perturbative regime.
The dot-dashed curve corresponds to the
extrapolated approximation given by Eq. (\ref{eq:alphafit}). Only the point-to-point uncorrelated
uncertainties  of the JLab data are used to estimate the uncertainties,  since a systematic shift cancels
in the derivative.
Nevertheless, the uncertainties are still large. 
The $\beta$-function extracted from LF holography, as well as the forms obtained from
the works of Cornwall~\cite{Cornwall:1981zr}, Bloch, Fisher {\it et al.},~\cite{S-Deq.} Burkert and Ioffe~\cite{Burkert-Ioffe} and Furui and Nakajima,~\cite{Furui}  are seen to have a similar shape
and magnitude.

Judging from these results, we infer that the   actual  $\beta$-function of QCD will extrapolate between the non-perturbative results for $Q < 1$ GeV and the pQCD results
for $Q > 1$ GeV. We also observe that the general conditions
\begin{eqnarray}
& \beta(Q \to 0) =  \beta(Q \to \infty) = 0 , \label{a} \\
&  \beta(Q)  <  0, ~ {\rm for} ~  Q > 0 , \label{b}\\
& \frac{d \beta}{d Q} \big \vert_{Q = Q_0}  = 0, \label{c} \\
& \frac{d \beta}{d Q}   < 0, ~ {\rm for} ~ Q < Q_0, ~~
 \frac{d \beta}{d Q}   > 0, ~ {\rm for} ~ Q > Q_0 \label{d} .
\end{eqnarray}
are satisfied by our model $\beta$-function obtained from LF holography.

Eq. (\ref{a}) expresses the fact that  QCD approaches a conformal theory in both the far ultraviolet and deep infrared regions. In the semiclassical approximation to QCD  without particle creation or absorption,
the $\beta$-function is zero and the approximate theory is scale  invariant
in the limit of massless quarks.~\cite{Parisi:1972zy} When quantum corrections are included,
the conformal behavior is
preserved at very large $Q$ because of asymptotic freedom and near $Q \to 0$ because the theory develops a fixed 
point.  An infrared fixed point is in fact a natural consequence of color confinement:~\cite{Cornwall:1981zr}
since the propagators of the colored fields have a maximum wavelength,  all loop
integrals in the computation of  the gluon self-energy decouple at $Q^2 \to 0$.~\cite{Brodsky:2008be} Condition (\ref{b}) for $Q^2$ large, expresses the basic anti-screening behavior of QCD where the strong coupling vanishes. The $\beta$-function in QCD is essentially negative, thus the coupling increases monotonically from the UV to the IR where it reaches its maximum value:  it has a finite value for a theory with a mass gap. Equation (\ref{c}) defines the transition region at $Q_0$ where the beta function has a minimum.  Since there is only one hadronic-partonic transition, the minimum is an absolute minimum; thus the additional conditions expressed in Eq (\ref{d}) follow immediately from 
Eqs.  (\ref{a}-\ref{c}). The conditions given by Eqs.  (\ref{a}-\ref{d}) describe the essential
behavior of the full $\beta$-function for an effective QCD coupling whose scheme/definition is similar to that of the $V$-scheme.

\section{Conclusions \label{conclusions}}

As we have shown, the combination of Anti-de Sitter  space
(AdS) methods with light-front (LF) holography provides a remarkably accurate first approximation for the spectra and wavefunctions of meson and baryon light-quark  bound states. We  obtain a connection between a semiclassical first approximation to QCD, quantized on the light-front, with hadronic modes propagating on a fixed AdS background. The resulting bound-state Hamiltonian equation of motion in QCD leads to  relativistic light-front wave equations in the invariant impact variable $\zeta$ which measures the separation of the quark and gluonic constituents within the hadron at equal light-front time. This corresponds 
to the effective single-variable relativistic Schr\"odinger-like equation in the AdS fifth dimension coordinate $z$,  Eq. (\ref{eq:QCDLFWE}). The eigenvalues give the hadronic spectrum, and the eigenmodes represent the probability distributions of the hadronic constituents at a given scale.
As we have shown, the light-front holographic mapping of  effective classical gravity in AdS space, modified by a positive-sign dilaton background,  provides a very good description of the spectrum and form factors of light mesons and baryons.

There are many phenomenological applications where detailed knowledge of the QCD coupling and the renormalized gluon propagator at relatively soft momentum transfer are essential.
This includes the rescattering (final-state and initial-state interactions) which
create the leading-twist Sivers single-spin correlations in
semi-inclusive deep inelastic scattering,~\cite{Brodsky:2002cx, Collins:2002kn} the Boer-Mulders functions which lead to anomalous  $\cos 2 \phi$  contributions to the lepton pair angular distribution in the unpolarized Drell-Yan reaction,~\cite{Boer:2002ju} and the Sommerfeld-Sakharov-Schwinger correction to heavy quark production at threshold.~\cite{Brodsky:1995ds}
The confining AdS/QCD coupling from light-front holography can lead to a
quantitative understanding of this factorization-breaking physics.~\cite{Collins:2007nk}

We have also shown that the light-front holographic mapping of  effective classical gravity in AdS space, modified by the same  positive-sign dilaton background  predicts the form of a non-perturbative effective coupling $\alpha_s^{AdS}(Q)$ and its $\beta$-function.
The AdS/QCD running coupling is in very good agreement with the effective
coupling $\alpha_{g_1}$ extracted from  the Bjorken sum rule. Surprisingly, the Furui and Nakajima lattice results~\cite{Furui} also agree better overall with the $g_1$ scheme rather than the $V$ scheme. Our analysis
indicates that light-front holography captures the essential dynamics of confinement. The holographic $\beta$-function displays a transition from  nonperturbative to perturbative  regimes  at a momentum scale $Q \sim 1$ GeV. It appears to captures  the essential characteristics of the full  $\beta$-function of QCD, thus giving further support to the application of the gauge/gravity duality to the confining dynamics of strongly coupled QCD.

\section*{Acknowledgments}

Presented by SJB at SCGT09, 2009 International Workshop on Strong Coupling Gauge Theories in the LHC Era,
Nagoya, December 8-11, 2009.
We thank  Volker Burkert, John Cornwall,   Sadataka Furui, 
Philipp H\"agler, Wolfgang Korsch, G. Peter Lepage, Takemichi Okui, Joannis Papavassiliou and Anatoly Radyushkin for helpful comments. 
This research was supported by the Department of Energy contracts DE--AC02--76SF00515 and  DE-AC05-84ER40150.

\end{document}